\documentclass[cameraready]{Interspeech}

\usepackage{booktabs}
\usepackage{graphicx}
\usepackage{subcaption}
\usepackage{tabularx}
\usepackage{soul}
\usepackage{hyperref}

\title{Speech Codec Probing from Semantic and Phonetic Perspectives}

\author[affiliation={1}, orcid=0000-0002-1387-5418, equalcontribution, correspondingauthor]{Xuan}{Shi}
\author[affiliation={}, equalcontribution]{Chang}{Zeng}
\author[affiliation={1},  orcid=0000-0002-2053-9068, equalcontribution]{Tiantian}{Feng}
\author[affiliation={1}, orcid=0009-0001-1351-847X]{Shih-Heng}{Wang}
\author[affiliation={2}, orcid=0000-0002-6765-9462]{Jianbo}{Ma}
\author[affiliation={1}, orcid=0000-0002-1052-6204]{Shrikanth}{Narayanan}

\address{
    $^1$ University of Southern California, USA \\
    $^2$ Dolby Laboratories, USA
}

\email{xuanshi@usc.edu}


\usepackage{comment}

\begin{document}

\maketitle

\begin{abstract}
Speech tokenizers are essential for connecting speech to large language models (LLMs) in multimodal systems. Speech tokenizers are expected to preserve both semantic and acoustic information for downstream understanding and generation tasks. However, emerging evidence suggests that the term "semantic" in speech processing does not align with linguistic lexical-semantic, leading to a mismatch between speech and text modality. In this paper, we systematically analyze the information encoded by several widely used speech tokenizers, evaluating their lexical-semantic and phonetic content through three tasks. Our results show that current tokenizers primarily capture phonetic rather than lexical-semantic structure, deriving practical implications for the design of next-generation speech tokenization methods. Code is released to public at \url{https://github.com/Alexuan/codec_probing_release}.
\end{abstract}

\section{Introduction}

The rapid advancement of large language models (LLMs) has driven significant interest in extending their sequence modeling capabilities to the speech modality. A growing body of research, including GPT-4o \cite{hurst2024gpt}, Qwen2.5-Omni \cite{xu2025qwen25omni}, and Moshi~\cite{defossez2024moshi}, demonstrates the potential of unified multimodal large language models (MLLMs) that can simultaneously understand and generate both text and speech. Central to these systems is the speech tokenizer, which converts continuous speech waveforms into discrete token sequences amenable to the autoregressive modeling paradigm of LLMs, enabling a shared vocabulary and modeling framework for diverse tasks spanning automatic speech recognition (ASR), spoken language understanding, text-to-speech synthesis, and voice conversion~\cite{coreteam2025mimoaudio,wang2024ham,du2024cosyvoice,li2025sonicsim}.

The design of speech tokenizers has gradually evolved from pure signal compression to semantic-aware encoding. Early neural audio codecs such as EnCodec \cite{defossez2022encodec} and DAC \cite{kumar2023dac} employ convolutional encoder-decoder architectures with Residual Vector Quantization (RVQ) \cite{gray1984vector, vasuki2006review}, optimizing reconstruction fidelity through perceptual and adversarial losses. While effective at preserving acoustic details, their representations are not explicitly designed to capture linguistic content \cite{defossez2022encodec, kumar2023dac}. To address this, AudioLM \cite{borsos2023audiolm} introduced a two-stage paradigm that separates ``semantic tokens'' derived from SSL models from ``acoustic tokens'' produced by neural codecs. This semantic-acoustic decomposition has since become widely adopted \cite{gong2025xy}, with MIMI \cite{defossez2024moshi} internalizing it by distilling WavLM \cite{chen2022wavlm} features into its first codebook layer. More recently, end-to-end systems such as Qwen2.5-Omni \cite{xu2025qwen25omni} and MiMo-Audio \cite{coreteam2025mimoaudio} directly jointly optimize speech tokenizer and LLM.

However, a critical question remains: does the information labeled as ``semantic” in speech tokens genuinely reflect linguistic meaning? Choi et al.~\cite{choi2024self} demonstrated that self-supervised learning (SSL) speech representations are more phonetic than semantic, exhibiting greater similarity between near-homophones (e.g., ``accept'' and ``except'') than between synonyms (e.g., ``big'' and ``large''). This suggests that the ``semantic tokens''may more accurately be characterized as phonetic representations. If this is the case, the misalignment between phonetic speech tokens and semantic text tokens may contribute to the performance degradation observed in MLLMs on speech understanding tasks \cite{xiang2025understanding,cuervo2025closing,zeng2023improving}. Therefore, probing the information encoded at each codebook layer is essential for understanding the algorithmic limitations and guiding future design.

In this paper, we systematically probe the knowledge encoded in speech codec representations from both semantic and phonetic perspectives. We first clarify the conceptual distinction between ``semantic'' and ``phonetic'' as used throughout our analysis in Sec. \ref{sec:problem_statement}. 
We then conduct three complementary probing experiments across four representative speech codecs, namely EnCodec \cite{defossez2022encodec}, DAC \cite{kumar2023dac}, MIMI \cite{defossez2024moshi}, and MIMO \cite{coreteam2025mimoaudio}, that span diverse architectural designs and training paradigms: (1) we extend an SSL-based probing analysis~\cite{choi2024self} to neural codecs to examine the semantic-phonetic knowledge encoded in their representation. Specifically, we conduct this probing analysis using synonym and near-homophone word pairs in Sec. \ref{sec:semantic_phonetic};
(2) We then introduce an articulatory phonetic probing analysis using Vocal Tract Distance (VTD) features \cite{kim2014enhanced, shi2024direct} extracted from real-time Magnetic Resonance Imaging (rt-MRI). This experiment provides a physiologically grounded assessment of phonetic encoding in Sec. \ref{sec:articulation_phonetic};
and (3) we finally evaluate the cross-modal semantic alignment between speech and text token spaces using Centered Kernel Alignment (CKA) in Sec. \ref{sec:semantic_alignment}. Through these analyses, we aim to provide actionable insights regarding the nature of information captured by current speech tokenizers, and to inform the design of future codecs that better serve language model integration. 

\section{Probing Experiment Settings}

In this section, we introduce the experiment settings for the probing experiments regarding semantic and phonetic distribution in speech codecs. 

\subsection{Probing Task: Semantic and Phonetic}
\label{sec:problem_statement}

Within the speech processing community, the term ``semantic'' is often used to describe representations derived from self-supervised speech models such as HuBERT~\cite{hsu2021hubert} and WavLM~\cite{chen2022wavlm}. However, this usage conflates two fundamentally distinct linguistic concepts.

In this paper, we adopt the following definitions. \textit{Semantic} refers to lexical meaning — the level at which synonyms such as ``big'' and ``large'' are considered close. \textit{Phonetic} refers to speech production — the level at which near-homophones such as ``accept'' and ``except'' are considered similar.

\subsection{Model Selection}
\label{sec:model_selection}

We select four representative speech codec models for our probing experiments: EnCodec \cite{defossez2022encodec}, DAC \cite{kumar2023dac}, MIMI \cite{defossez2024moshi}, and MIMO \cite{coreteam2025mimoaudio}. These models span a diverse range of architectural designs, training strategies, and intended applications, enabling a comprehensive analysis of how different codec design choices affect the amount of semantic and phonetic information preserved in the learned discrete units.
Table \ref{tab:speech_tokenizers} summarizes the key design choices of each model.

\noindent\textbf{EnCodec} \cite{defossez2022encodec} employs a convolutional encoder-decoder architecture (SEANet \cite{tagliasacchi2020seanet}) with RVQ \cite{gray1984vector, vasuki2006review}. 
It is trained end-to-end with a combination of reconstruction, perceptual, and adversarial losses.
As one of the earliest and most widely adopted neural audio codecs, EnCodec is a pure compression model without explicit semantic or phonetic modeling.

\noindent\textbf{DAC} \cite{kumar2023dac} builds upon the RVQ-based codec framework with several architectural improvements, including factorized and L2-normalized codebook vectors \cite{yu2021vector} for better codebook utilization, snake activation functions \cite{ziyin2020neural} for improved periodic signal modeling, and quantizer dropout during training. 

\noindent\textbf{MIMI} \cite{defossez2024moshi} is the codec component of the Moshi real-time dialogue system. Its key distinguishing feature is a ``semantic''-acoustic disentangled codebook design: the first quantizer layer is distilled from WavLM \cite{chen2022wavlm}, a SSL model, to capture higher-level ``semantic'' representations, while the remaining layers encode residual acoustic information through RVQ. 

\noindent\textbf{MIMO-Audio-Tokenizer} \cite{coreteam2025mimoaudio} (hereafter referred to as ``MIMO'') is a Transformer-based codec developed within the MiMo-Audio framework. Unlike the codecs above, MIMO is trained from scratch through a two-stage process: joint audio reconstruction and ASR alignment with a LLM, followed by adversarial fine-tuning. 

\begin{table*}[t]
    \centering
    \caption{Comparison of different speech tokenizers across various dimensions. SR denotes sampling rate. Input and Output SR show the configuration we used in probing experiments.}
    \label{tab:speech_tokenizers}
    \begin{tabularx}{\textwidth}{lllXX} 
        \toprule
        \textbf{Speech Tokenizer} & \textbf{Input SR (Hz)} & \textbf{Output SR (bps)} & \textbf{Training Objectives} & \textbf{Application Tasks} \\ 
        \midrule
        EnCodec \cite{defossez2022encodec} & 24k & 12k & Reconstruction & Audio Compression \\ 
        DAC \cite{kumar2023dac} & 24k & 24k & Reconstruction & Audio Compression \\ 
        MIMI \cite{defossez2024moshi} & 24k & 4.4k & Reconstruction and Distillation & Conversational AI \\ 
        MIMO \cite{coreteam2025mimoaudio} & 24k & 1.55k & Reconstruction and ASR & Conversational AI \\ 
        \bottomrule
    \end{tabularx}
\vspace{-5mm}
\end{table*}

\subsection{Probing Semantic and Phonetic Knowledge in Codecs}
\label{sec:semantic_phonetic}

In this section, we investigate how much semantic and phonetic knowledge is preserved in different speech codebooks. 
We adopt the experimental design from a prior probing study on SSL speech models \cite{choi2024self}, motivated by the structural similarities between SSL models and speech codecs. 
Following \cite{choi2024self}, semantic and phonetic information density can be estimated via a proxy task that measures the feature distance between synonym and near-homophone word pairs. Specifically, a larger feature distance between synonym pairs relative to near-homophone pairs indicates a relative higher concentration of phonetic information, and vice versa.

\noindent \textbf{Data Construction}
We follow the methodology of \cite{choi2024self} to construct synonym and near-homophone word pairs. 
Individual word segments are extracted from the LibriSpeech \cite{panayotov2015librispeech} dataset using alignment timestamps generated by the Montreal Forced Aligner (MFA) \cite{mcauliffe2017montreal}. 
For each word, synonyms are retrieved from the cognitive synonym sets in WordNet \cite{miller1995wordnet}, a large-scale English lexical database. 
Regarding near-homophones, all words are first converted into phonemes using the CMU Pronouncing Dictionary \cite{cmudict}. We then calculate the phonetic similarity between words using the Levenshtein distance \cite{levenshtein1966binary}; word pairs with a normalized Levenshtein distance of less than 0.4 are classified as near-homophone pairs \cite{choi2024self}. 
While this study primarily utilizes English datasets due to the availability of well-characterized resources, we note that prior work in \cite{choi2024self} has demonstrated consistent trends in cross-lingual scenarios. 

\noindent \textbf{Feature Extraction}
We extract the accumulated decoded speech features from each codec layer. Specifically, the feature at a given layer is defined as the summation of the decoded output from the current quantizer layer and the accumulated features from all preceding layers. 
Since many speech codecs use Euclidean Residual Vector Quantizer, we employ the Euclidean distance as a standardized metric to compute the distances between these high-dimensional feature vectors. For more technical implementation details, we refer the reader to \cite{choi2024self}.

\vspace{-2mm}
\subsection{Articulatory Phonetic Probing}
\label{sec:articulation_phonetic}

\begin{figure}
    \centering
    \begin{subfigure}{0.23\textwidth}
        \includegraphics[width=\textwidth]{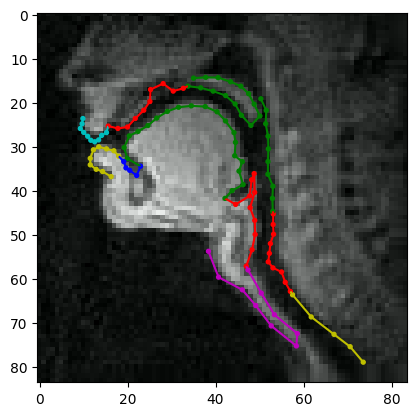}
        \caption{Articulator Boundary}
    \end{subfigure}
    \hfill
    \begin{subfigure}{0.23\textwidth}
        \includegraphics[width=\textwidth]{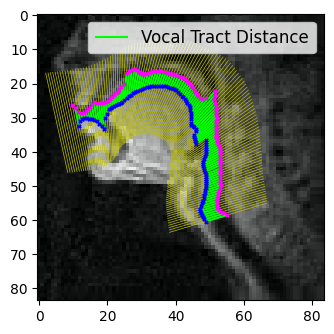}
        \caption{Vocal Tract Distance}
    \end{subfigure}
    \caption{Visualization of vocal tract mid-sagittal rt-MRI frames and corresponding VTD annotations.}
    \label{fig:articulation}
    \vspace{-5mm}
\end{figure}

To further quantitatively investigate the phonetic information encoded within speech codecs, we examine the correlation between articulatory dynamics and speech codec features. Specifically, we extract an articulatory feature, namely Vocal Tract Distance (VTD) \cite{kim2014enhanced, shi2024direct}, from the mid-sagittal plane of rt-MRI sequences. As illustrated in Fig. \ref{fig:articulation}, VTD represents the length of the intersections between the articulator boundaries and a set of gridlines \cite{kim2014enhanced}. These gridlines are evenly distributed from the lips to the larynx and are oriented perpendicular to the vocal tract midline. VTD effectively captures the shape of the airway tunnel during speech, providing a straightforward and comprehensive representation of phonetic information.

\noindent \textbf{Dataset}
Our articulatory analysis is conducted on the 75-Speaker dataset \cite{lim2021multispeaker} and the 75-Speaker Annot-16 subset \cite{shi25g_interspeech} (hereafter referred to as ``Annot-16''). The 75-Speaker dataset is a multimodal corpus containing simultaneous speech and rt-MRI video. This American English corpus comprises various speech materials, including vowel-consonant sequences, scripted reading, and spontaneous speech. Annot-16 provides expert-labeled articulator boundary annotations for 16 speakers from the original dataset.

\noindent \textbf{Evaluation Metrics}
We employ Projection Weighted Canonical Correlation Analysis (PWCCA) \cite{morcos2018insights} to measure the correlation between VTD and speech codec features. The VTD features are sampled at 83 Hz, matching the rt-MRI frame rate. For each speaker, the number of gridlines is calibrated to 120, where each index corresponds to a specific articulatory location. Consequently, the VTD is a time-sequence feature with a dimensionality of 120. To address the discrepancy in sampling rates and dimensions between VTD and codec features, we up-sample the VTD sequences to align with the codec features, ensuring a robust PWCCA calculation.

\subsection{Semantic Alignment between Text and Speech} 
\label{sec:semantic_alignment}

One of the primary applications of speech codecs is the conversion of raw audio into discrete speech tokens for use in speech language models (SLMs). Inspired by \cite{huh2024position}, we assume that the latent representation from text and speech, even though they are derived from different modalities, tend to move to a common representation space during joint training.
Motivated by this observation, we extract the latent representations that are decoded from text and speech tokens before LLM processing and measure the similarities between corresponding text-speech pairs in the shared representation space.
In this experiment, we choose MIMI \cite{defossez2024moshi} and MIMO \cite{coreteam2025mimoaudio} models for measuring representation similarities as they are originally designed for conversational AI and their compatible LLMs are accessible.


\noindent \textbf{Dataset}
We follow the word-level speech extraction pipeline described in Sec. \ref{sec:semantic_phonetic} to extract speech segments from the LibriSpeech dataset \cite{panayotov2015librispeech}. 

\noindent \textbf{Evaluation Metrics}
While alignment between text and speech is desirable, the latent feature spaces for different modalities in multimodal large language models (MLLMs) are not necessarily shared, which poses a significant challenge for direct evaluation. Therefore, we adopt the  Centered Kernel Alignment (CKA) \cite{kornblith2019similarity} to quantify the structural similarity between the speech and text modalities.

\section{Probing Experiment Findings}

\begin{figure*}[t] 
    \centering
    \begin{subfigure}{0.21\textwidth}
        \includegraphics[width=\textwidth]{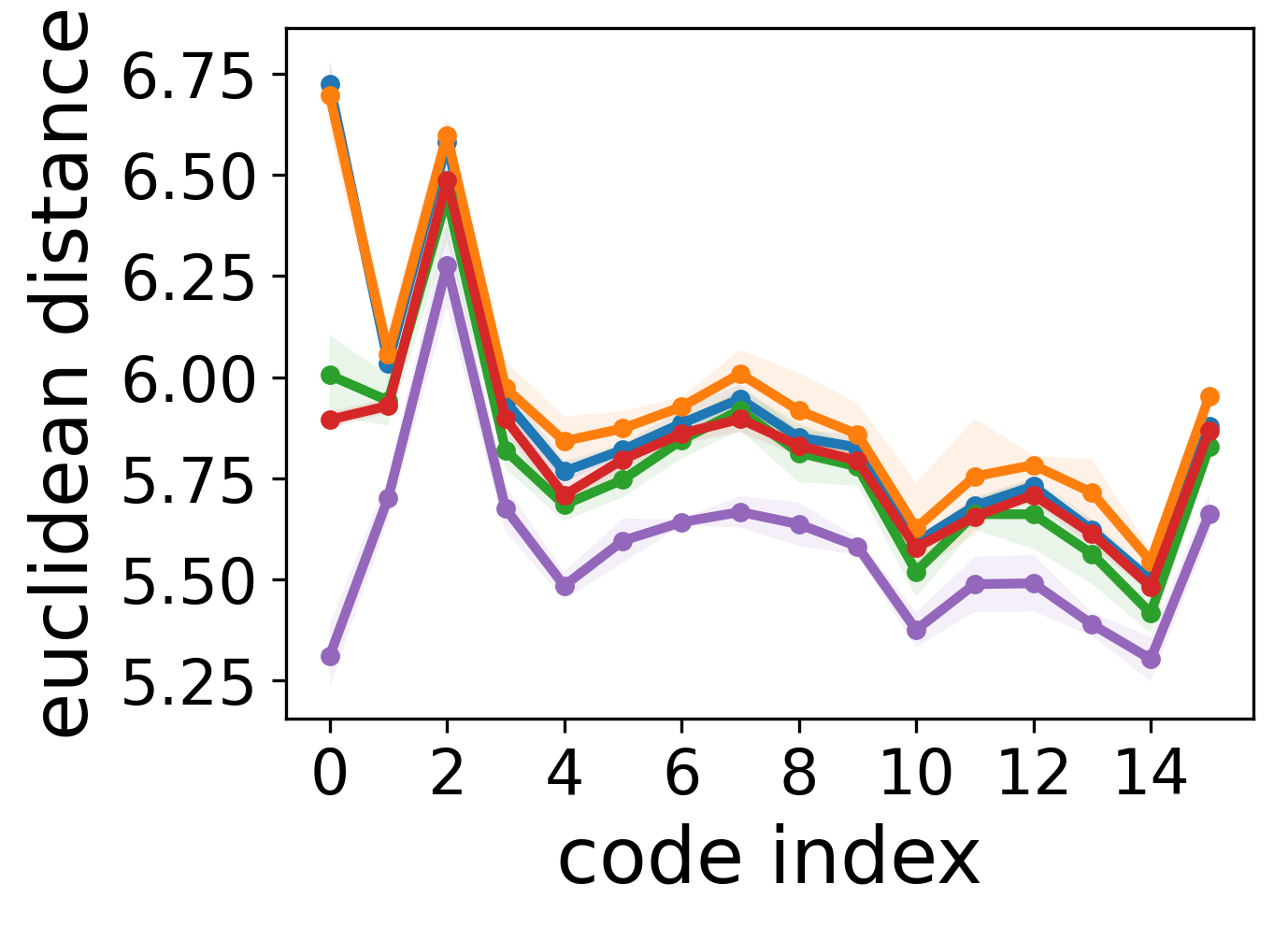}
        \caption{EnCodec}
    \end{subfigure}
    \hfill
    \begin{subfigure}{0.2\textwidth}
        \includegraphics[width=\textwidth]{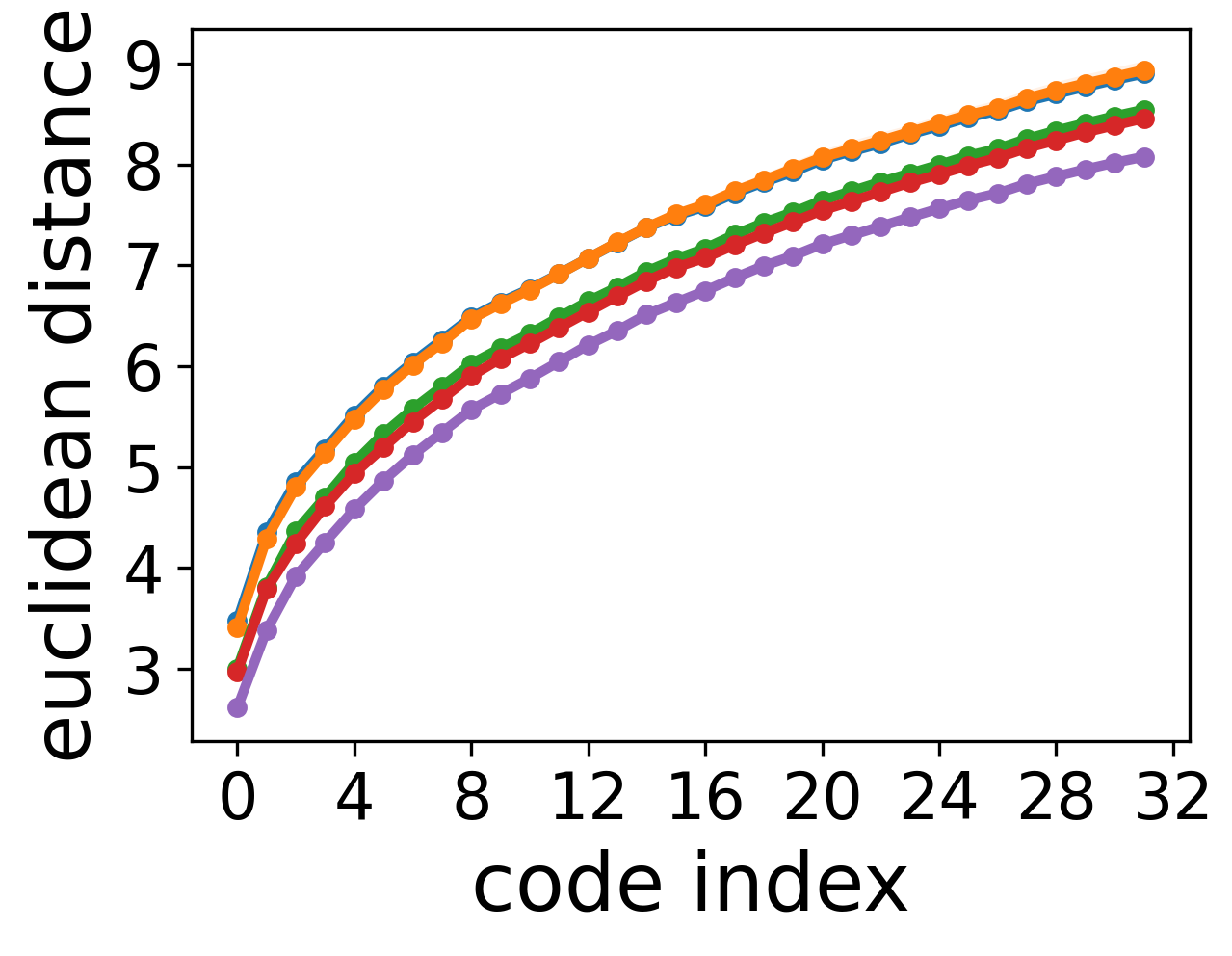}
        \caption{DAC}
    \end{subfigure}
    \hfill
    \begin{subfigure}{0.225\textwidth}
        \includegraphics[width=\textwidth]{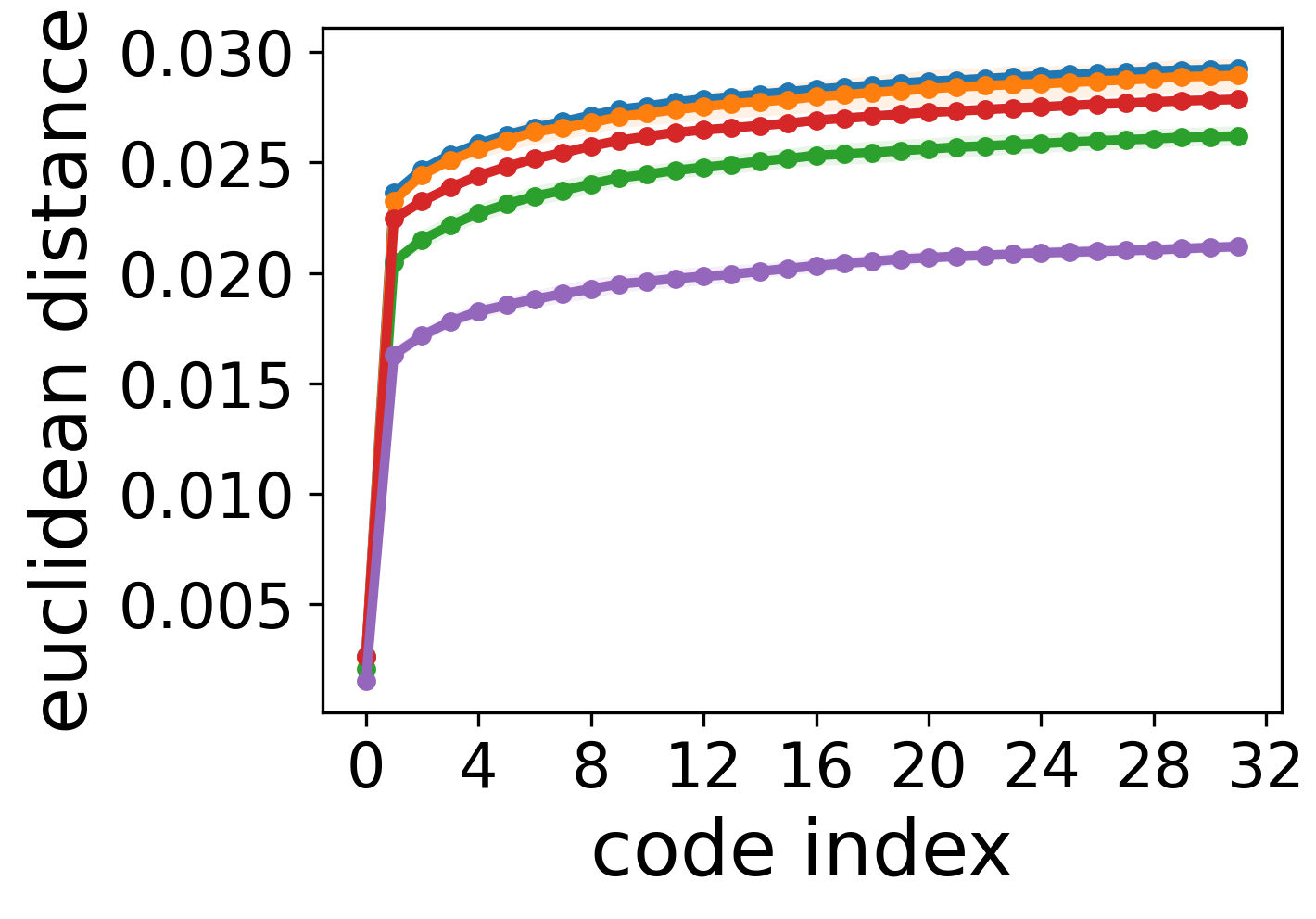}
        \caption{MIMI}
    \end{subfigure}
    \hfill
    \begin{subfigure}{0.31\textwidth}
        \includegraphics[width=\textwidth]{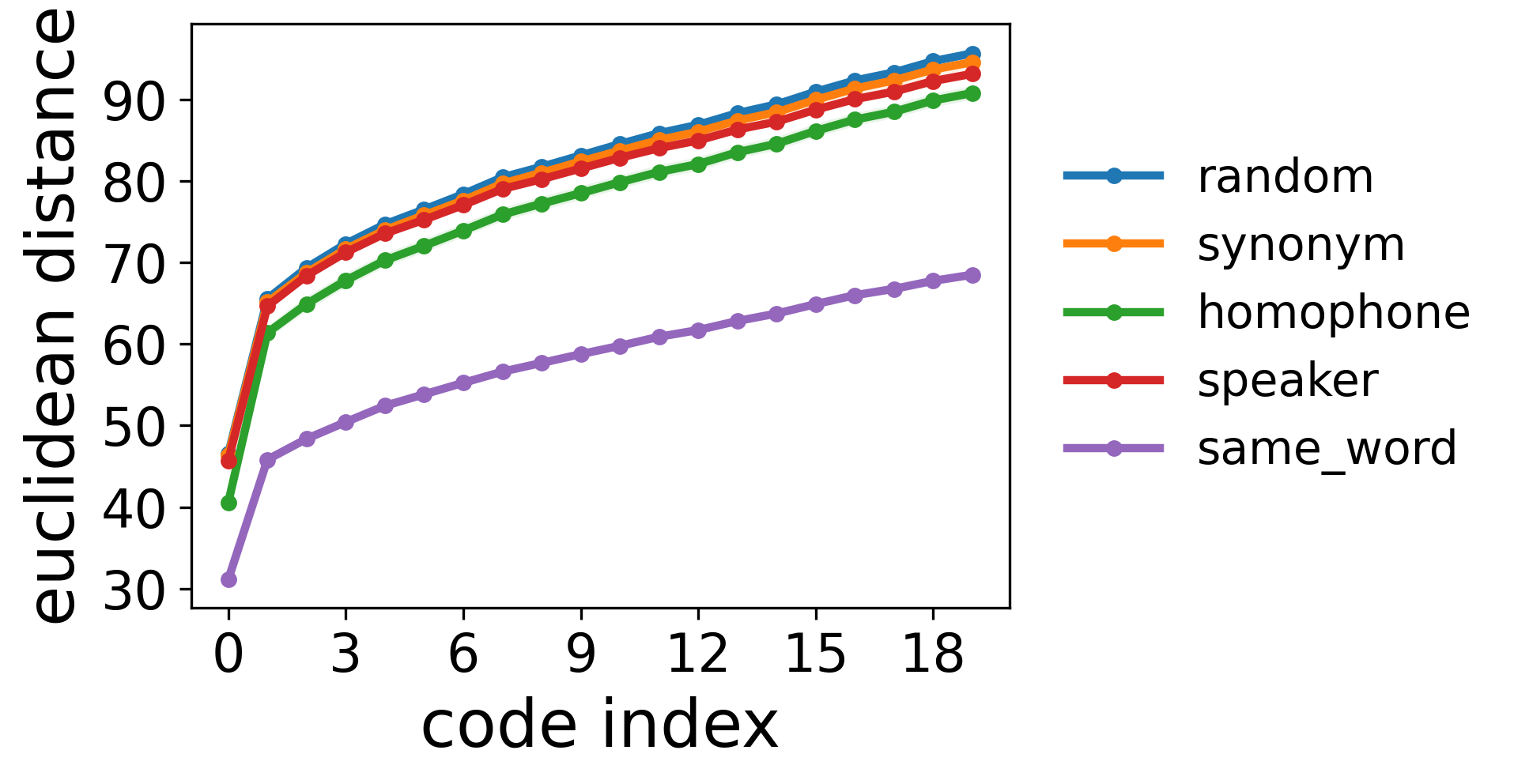}
        \caption{MIMO}
    \end{subfigure}

    \begin{subfigure}{0.22\textwidth}
        \includegraphics[width=\textwidth]{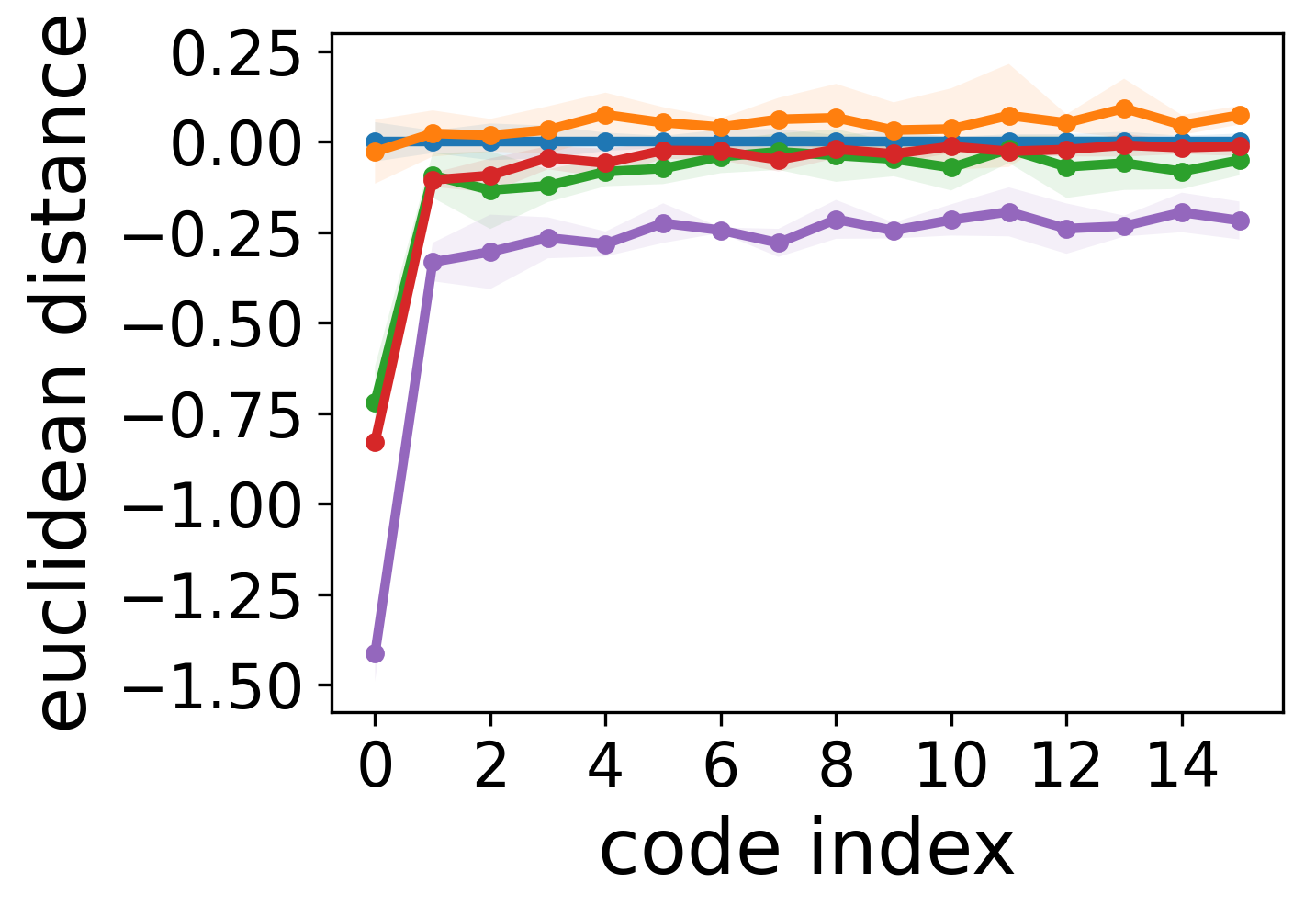}
        \caption{EnCodec - Norm}
    \end{subfigure}
    \hfill
   \begin{subfigure}{0.21\textwidth}
        \includegraphics[width=\textwidth]{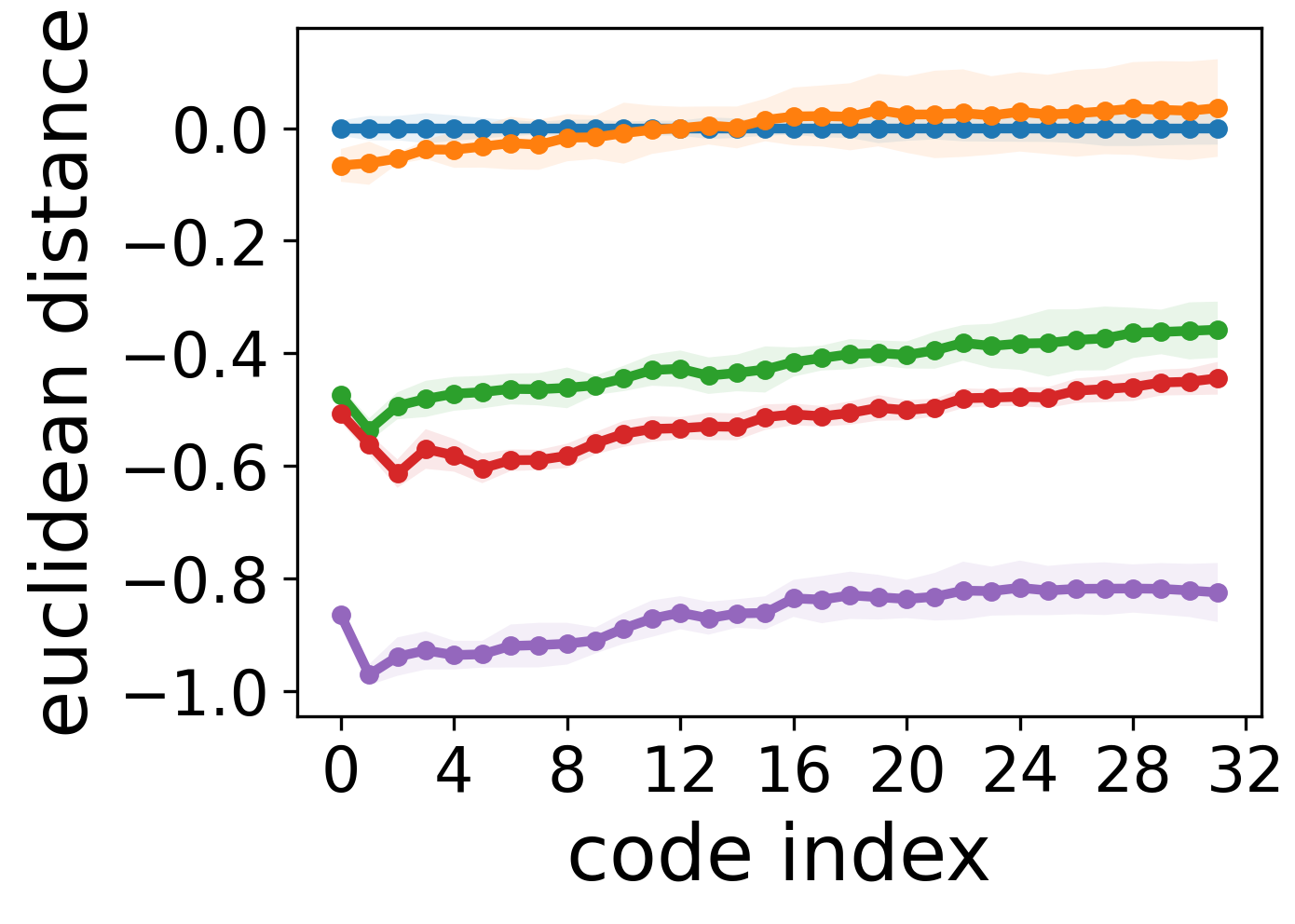}
        \caption{DAC - Norm}
    \end{subfigure}
    \hfill
    \begin{subfigure}{0.225\textwidth}
        \includegraphics[width=\textwidth]{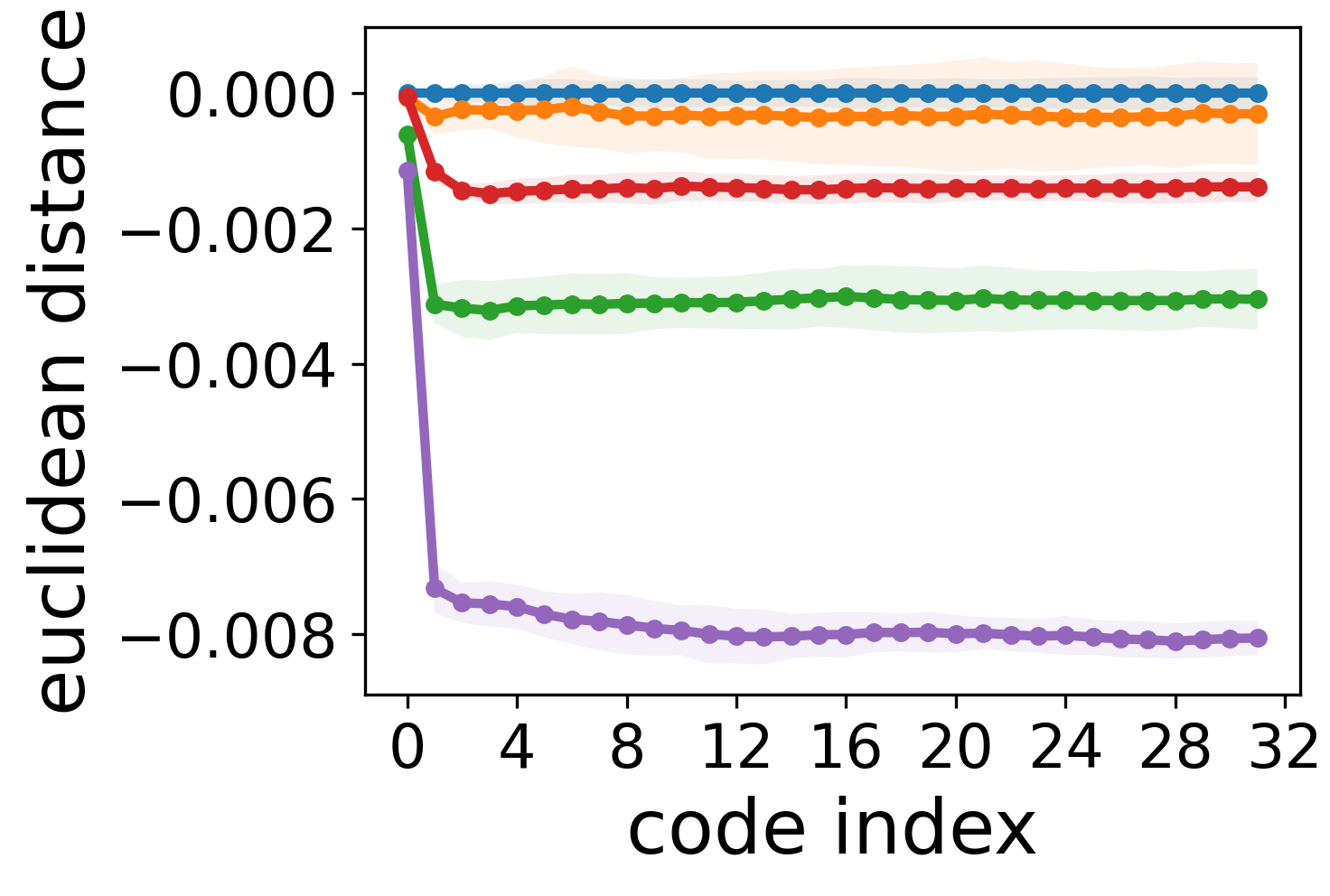}
        \caption{MIMI - Norm}
    \end{subfigure}
    \hfill
    \begin{subfigure}{0.3\textwidth}
        \includegraphics[width=\textwidth]{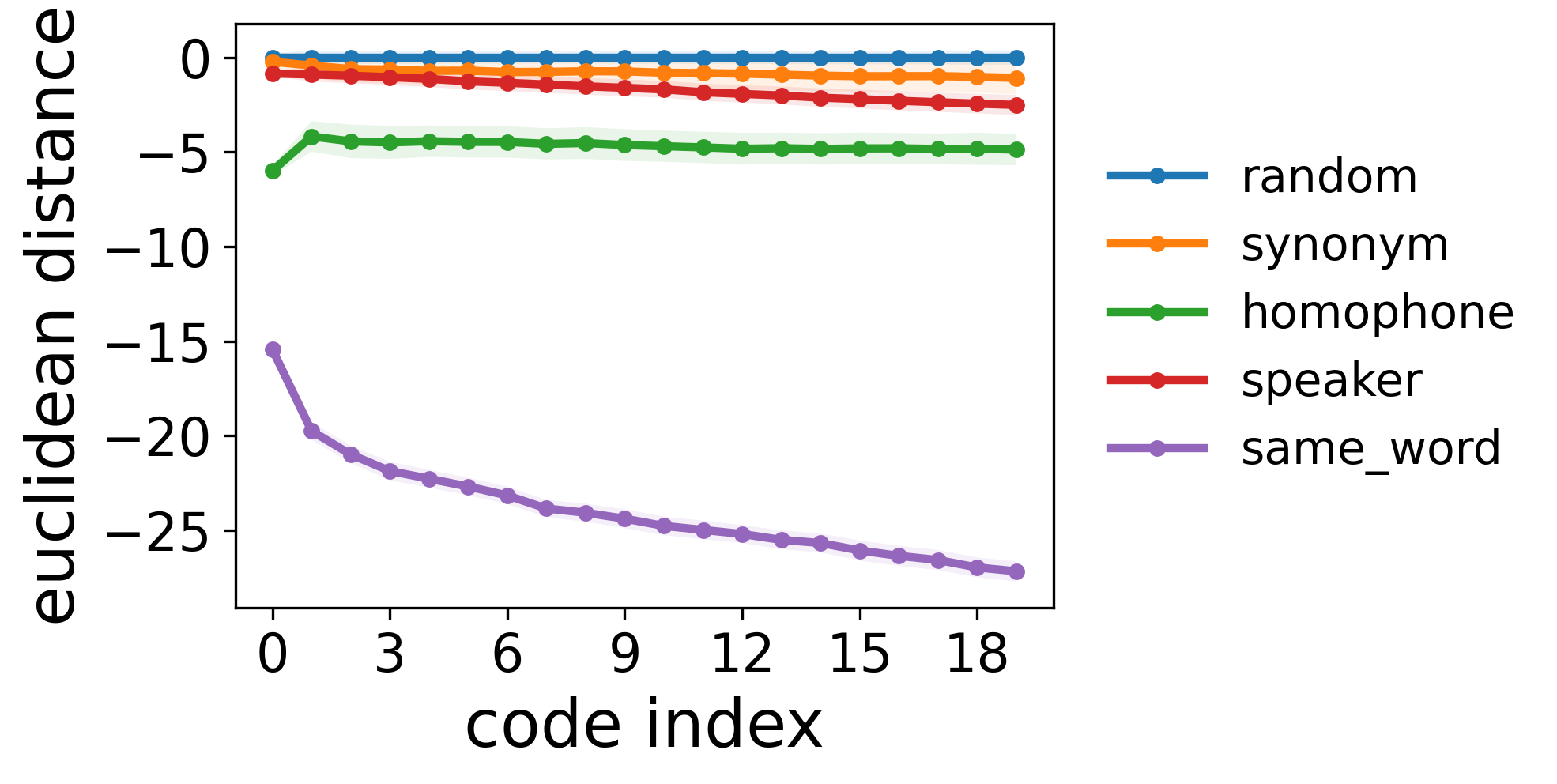}
        \caption{MIMO - Norm}
    \end{subfigure}
    \caption{Phonetic and Semantic Information Distribution}
    \label{fig:semantic_phonetic}

\end{figure*}

\begin{figure*}[t]
    \begin{subfigure}{0.24\textwidth}
        \includegraphics[width=\textwidth]{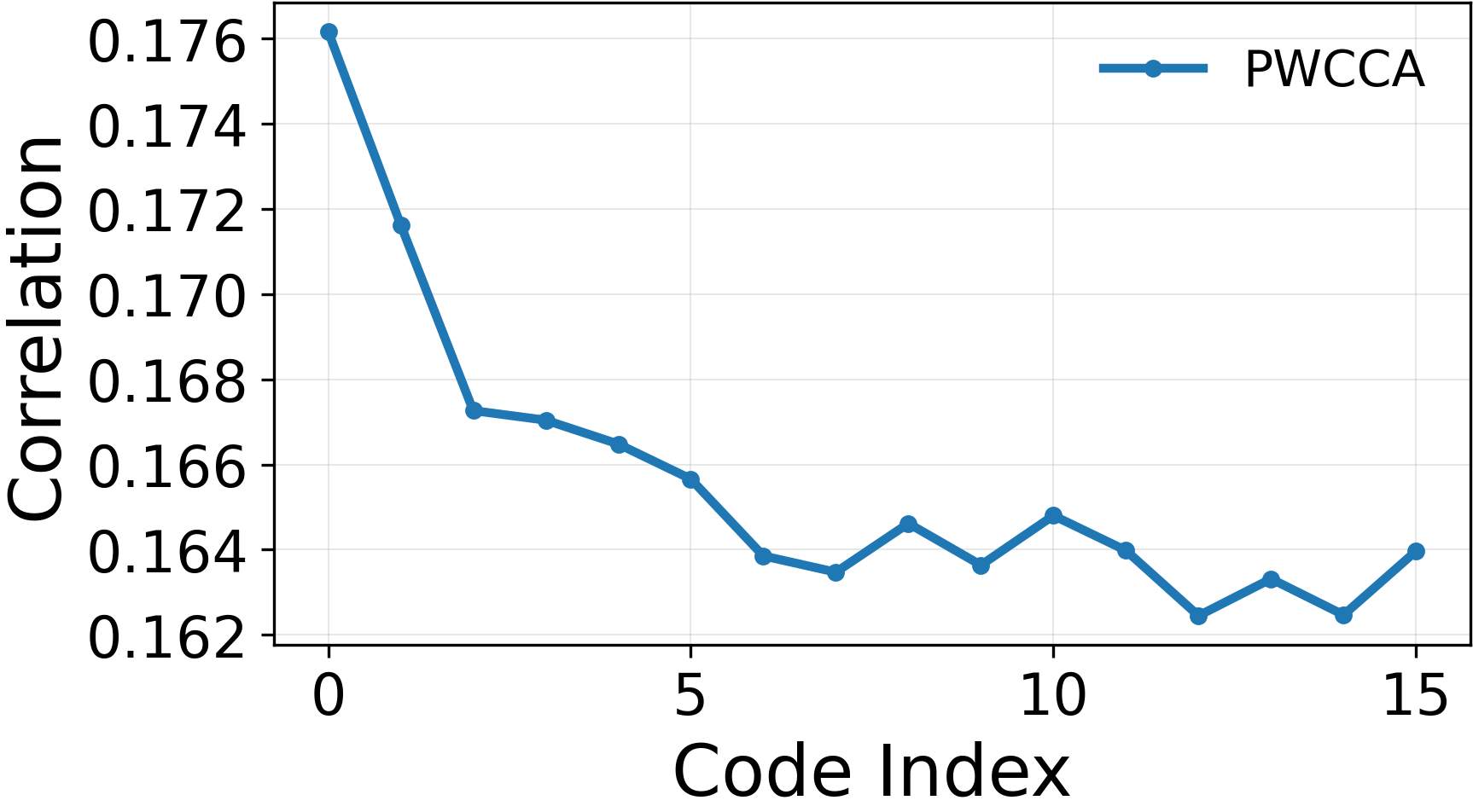}
        \caption{EnCodec}
    \end{subfigure}
    \hfill
    \begin{subfigure}{0.24\textwidth}
        \includegraphics[width=\textwidth]{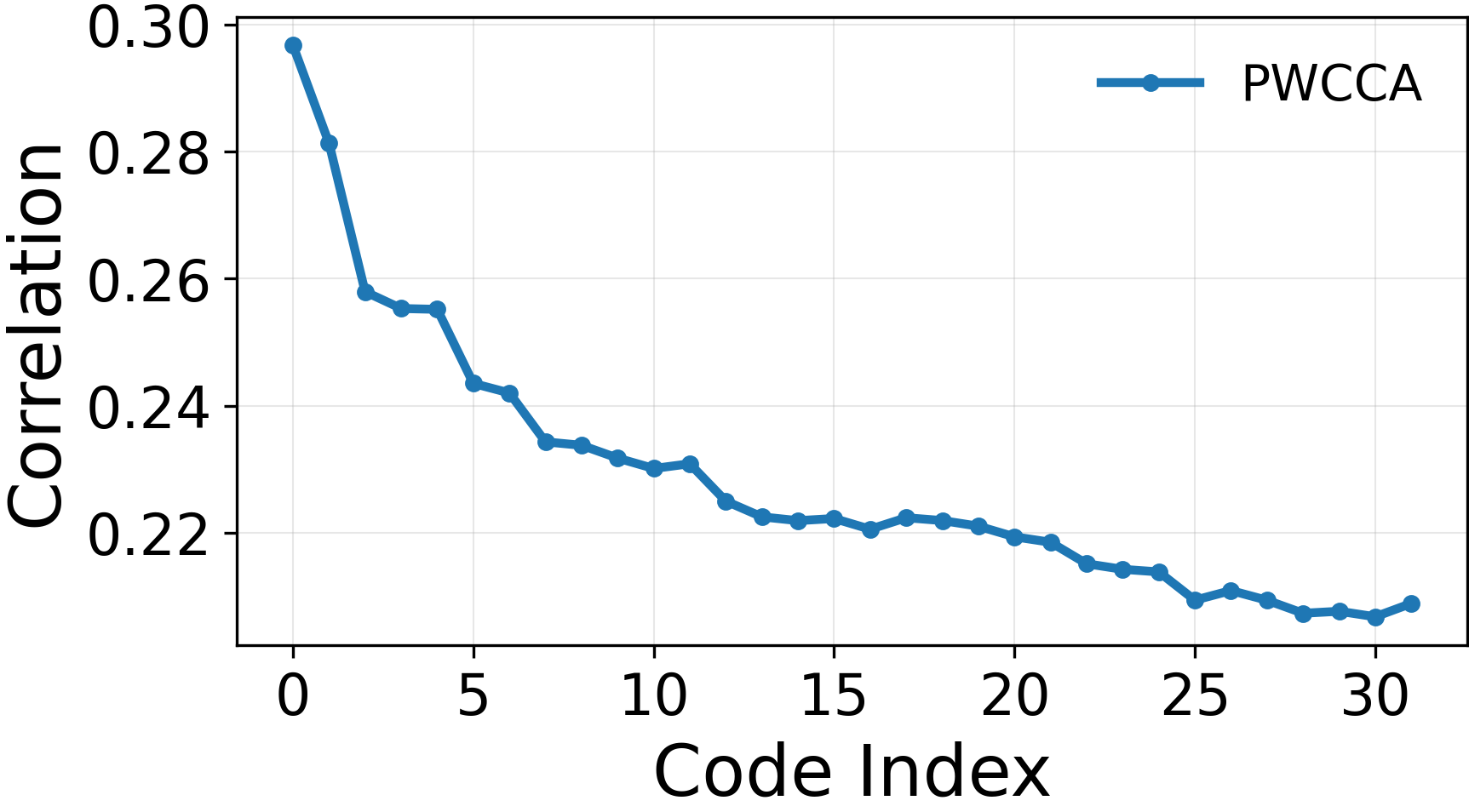}
        \caption{DAC}
    \end{subfigure}
    \hfill
    \begin{subfigure}{0.24\textwidth}
        \includegraphics[width=\textwidth]{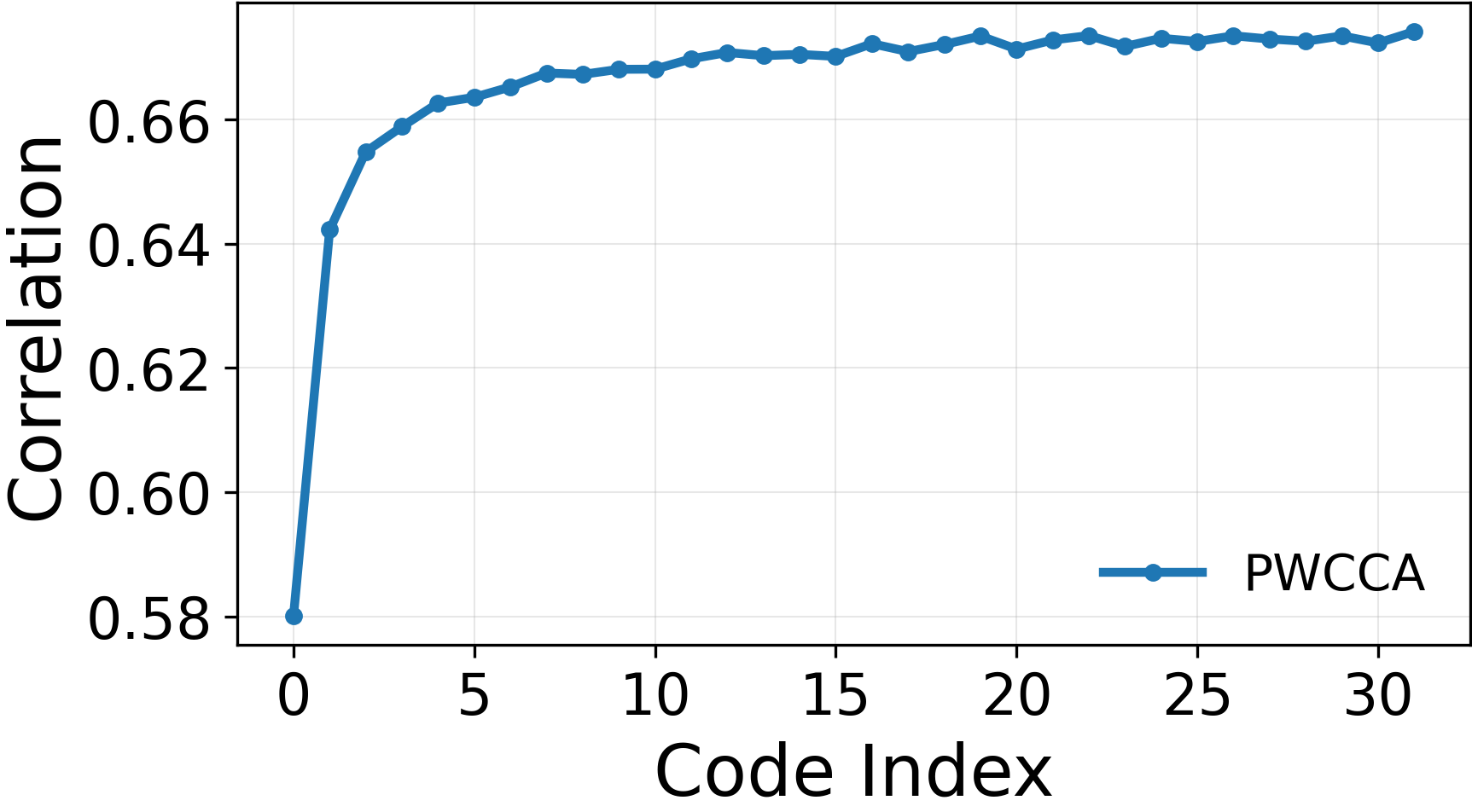}
        \caption{MIMI}
        \label{fig:mimi_phonetic}
    \end{subfigure}
    \hfill
    \begin{subfigure}{0.24\textwidth}
        \includegraphics[width=\textwidth]{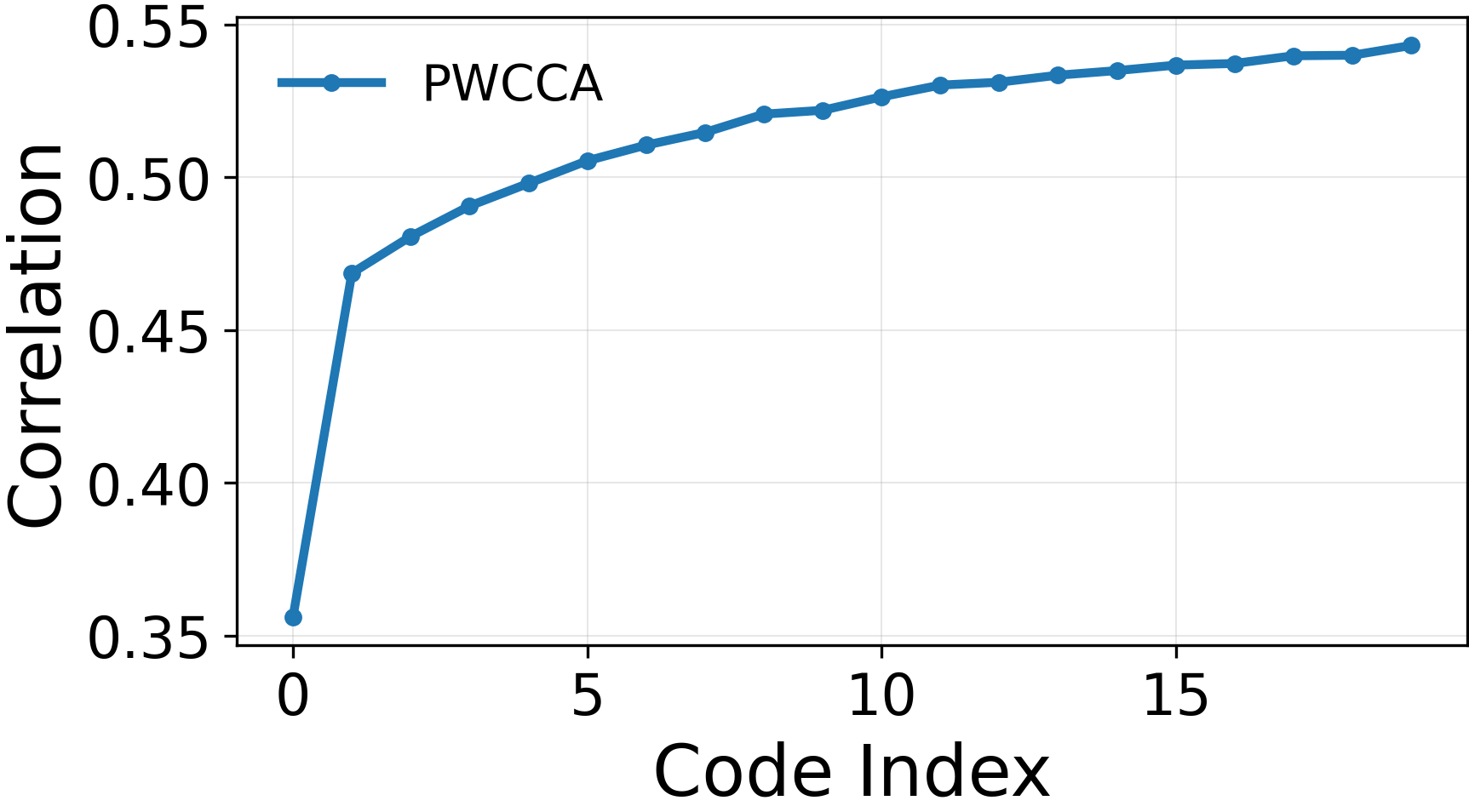}
        \caption{MIMO}
    \end{subfigure}
    \vspace{-1mm}
    \caption{Articulation Phonetic Correlation}
    \label{fig:phonetic_corr}
    \vspace{-3mm}
\end{figure*}

\subsection{Semantic and Phonetic Analysis}

We use the Euclidean distance between speech features extracted from word pairs as a proxy to evaluate how different codecs capture target semantic and phonetic information. As illustrated in Fig. \ref{fig:semantic_phonetic}, the top and bottom rows present the absolute Euclidean distances and the normalized feature distances, respectively. The normalized feature distance is calculated by subtracting the distances of the random setting.

From the absolute distance plots, we observe that as the codebook depth increases, EnCodec fluctuates without a clear pattern across all word-pair settings. In contrast, the remaining speech codecs exhibit a consistent increase in Euclidean distance as the codebook layer increases, indicating a progressive accumulation of information.

The normalized distributions in the bottom row provide a clearer perspective on the relative information density of semantic versus phonetic attributes. 

As the codebooks go deeper, most curves approach the random line. This trend suggests a "fading" effect of distinguishable semantic and phonetic information within the speech tokens.
From EnCodec and DAC, we observe the fading in a drastic and gradual pattern, respectively.
This fading is particularly evident for semantic information: the synonym curves for both EnCodec and DAC frequently overlap with or even exceed the random baseline starting from the early-to-mid codebook stages.
Notably, DAC maintains significantly higher distances for speaker attributes compared to synonyms and homophones, underscoring its superior modeling capability in capturing timbre and fine-grained acoustic details.
We attribute the fading effect in EnCodec and DAC, especially regarding semantic information, to the training objectives of these codec models, as they did not incorporate semantic-related tasks during training.

Both MIMI and MIMO demonstrate similar architectural trends: semantic, phonetic, and speaker-related information accumulate through the codebook layers. However, MIMI exhibits an earlier convergence of information and contains a higher proportion of phonetic (homophone) and acoustic (speaker) information. This behavior is likely attributed to MIMI’s first codebook being distilled from WavLM, which injects robust phonetic priors.

Overall, the evaluation across all four speech codecs consistently reveals that they preserve a substantially higher proportion of phonetic information relative to semantic information.

The above analysis establishes, from a functional perspective, that speech codecs consistently encode more phonetic than semantic information. However, this finding is derived from an indirect proxy task based on word-pair distances in the feature space. A natural question arises: does this phonetic dominance reflect a genuine encoding of speech production mechanisms, or is it merely an artifact of acoustic similarity? To answer this, we turn to articulatory evidence in the following section.

\vspace{-2mm}
\subsection{Articulation Phonetic Analysis}
\label{sec:articulation_phonetic_analysis}

To provide physiological grounding for the phonetic dominance observed in Sec. 4.1, we examine the correlation between codec features and articulatory dynamics measured by VTD, as shown in Fig.~\ref{fig:phonetic_corr}.
Cross validated with Fig~\ref{fig:semantic_phonetic}, EnCodec and DAC demonstrate the fading of phonetic information as the correlation curve keep go down in Fig.~\ref{fig:phonetic_corr}. Likewise, MIMI and MIMO show the phonetic accumulation with the codebooks go deeper. The early converge point also appear in the phonetic correlation from MIMI and MIMO. 

\begin{figure}[ht]
    \centering
    \includegraphics[width=0.5\linewidth]{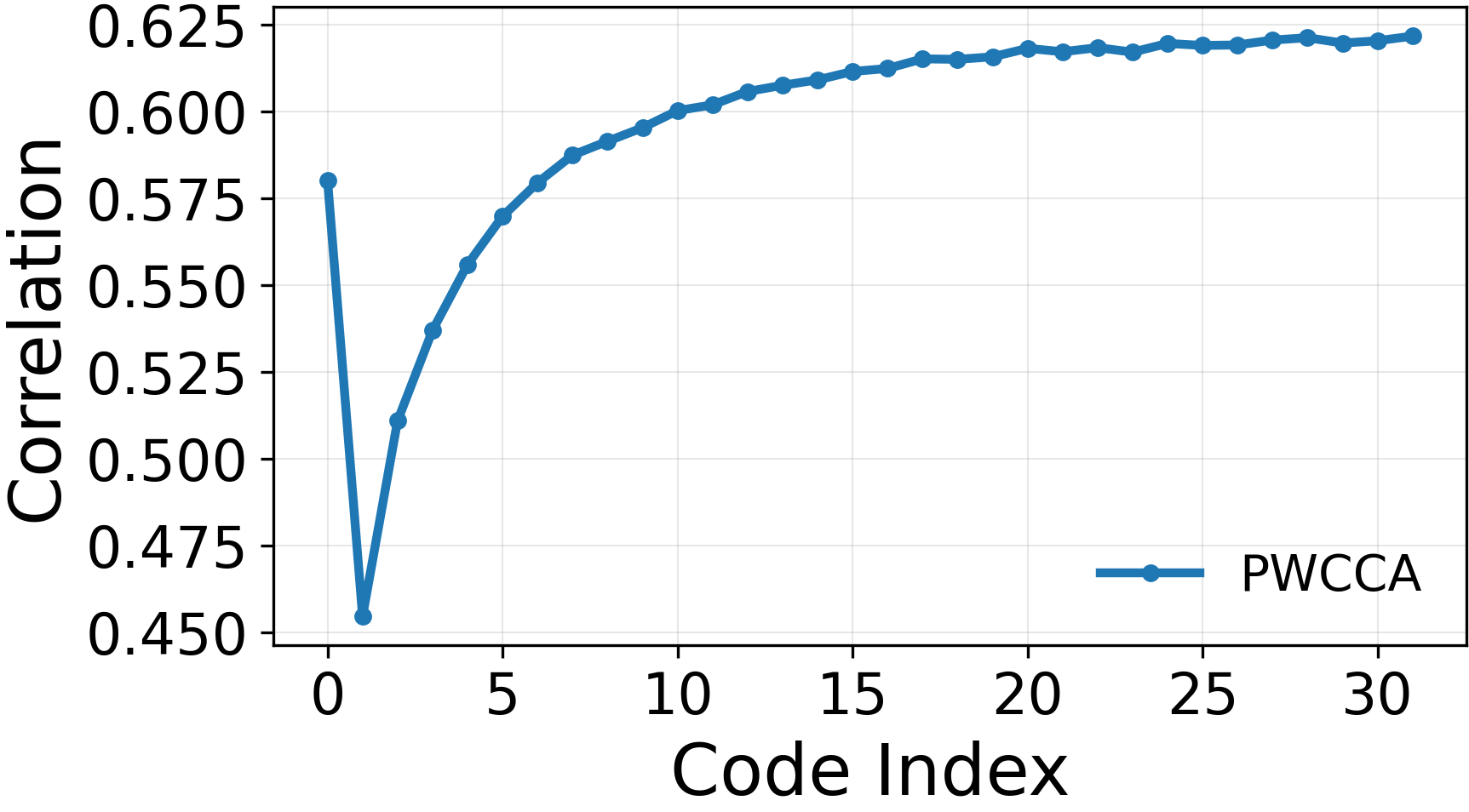}
    \vspace{-1mm}
    \caption{MIMI - Separate Phonetic Analysis}
    \label{fig:mimi_phonetic_separate}
    \vspace{-2mm}
\end{figure}

\noindent \textbf{Phonetic Analysis for MIMI}
In the MIMI architecture \cite{defossez2024moshi}, the feature from the first codebook layer, commonly referred to as the ``semantic'' feature, is distilled from WavLM \cite{chen2022wavlm}. However, since WavLM is primarily trained on ASR and acoustic-related tasks, its representations may not directly align with the linguistic ``semantic'' concept introduced in Sec. \ref{sec:problem_statement}. To investigate this, we separately calculated the phonetic correlation between the first-layer feature and the accumulated features of the remaining layers using VTD. A comparison between Fig. \ref{fig:mimi_phonetic_separate} and Fig. \ref{fig:mimi_phonetic} clearly reveals that distillation from WavLM injects significant phonetic-related knowledge. Furthermore, even without the first layer, as the acoustic codebook layers go deeper, the speech features from these layers also accumulate a substantial portion of phonetic information. 


\subsection{Semantic Alignment}

The preceding analyses consistently show that codec tokens lack semantic information. A direct consequence would be weak structural alignment between speech codec token and text token. To test this, we measure Centered Kernel Alignment~(CKA) between the two spaces. Both MIMI and MIMO yield CKA scores well below 1.0 (0.329 and 0.122, respectively), confirming a substantial cross-modal structural gap. To control for spurious alignment caused by each space's intrinsic geometry, we also compute a random-permutation baseline that breaks the speech--text pairing. Both codecs show only modest gains over this baseline ($\Delta{=}0.087$ for MIMI; $\Delta{=}0.054$ for MIMO), indicating that neither codec captures strong semantic structure—consistent with Fig.~\ref{fig:semantic_phonetic}. MIMI's higher absolute CKA is likely an artifact of the low effective dimensionality inherited from its WavLM-distilled first layer, which inflates chance alignment; the baseline-corrected $\Delta$ provides a fairer comparison.

\subsection{Discussion}
\label{sec:discussion}
Across three complementary analyses, our results paint a consistent picture: speech codecs preserve substantially more phonetic than semantic information, this phonetic encoding is grounded in articulatory production mechanisms, and the resulting representations show weak structural alignment with text semantics. These findings suggest two potential directions for improving semantic encoding in speech tokenizers.
First, since distillation from SSL speech models inherits their phonetic bias (as demonstrated in Sec. \ref{sec:articulation_phonetic_analysis}), future tokenizers may benefit from distilling representations from models with genuine text semantic understanding, such as the text embeddings of LLMs or cross-modal encoders trained on paired speech-text data. Second, the training objective itself could be augmented with explicit semantic constraints, for instance by encouraging codec representations of synonymous words to be closer in the latent space, thereby complementing acoustic reconstruction with semantic structure preservation.
\section{Conclusion}

This paper systematically probes the information encoded in speech codec representations from semantic and phonetic perspectives. Through functional analysis, articulatory probing with rt-MRI data, and cross-modal alignment evaluation, applied across four representative codec frameworks, we demonstrate that current speech codecs consistently encode more phonetic than semantic information. The articulatory evidence confirms that this phonetic dominance reflects genuine speech production mechanisms. Our analysis of MIMI further reveals that distillation from WavLM injects phonetic rather than semantic knowledge, challenging the widespread use of the term ``semantic tokens.'' These findings highlight the need for future work on tokenizer designs that incorporate explicit semantic objectives to better serve language model integration.





\section{Acknowledgement}
The work was supported in part by NSF (IIS 2311676), IARPA ARTS, and Dolby.

\section{Generative AI Use Disclosure}
We used Gemini 3.1 and Claude Opus 4.6 for paper proofreading. All substantive contents remain the authors’ own.

\bibliographystyle{IEEEtran}
\bibliography{mybib}

\end{document}